\documentclass[preprint,showpacs,preprintnumbers,amsmath,amssymb]{revtex4}

\usepackage{graphicx}
\usepackage{dcolumn}
\usepackage{bm}

\begin{document}


\title{Special features of the $^9$Be$\rightarrow$2He fragmentation in emulsion 
at an energy of 1.2~A~GeV}

\author{D.~A.~Artemenkov}
       \email{artemenkov@lhe.jinr.ru}  
   \affiliation{Joint Insitute for Nuclear Research, Dubna, Russia}
 \author{V.~Bradnova}
   \affiliation{Joint Insitute for Nuclear Research, Dubna, Russia} 
\author{M.~M.~Chernyavsky}
  \affiliation{Lebedev Institute of Physics, Russian Academy of Sciences, Moscow, Russia} 
\author{N.~A.~Kachalova}
   \affiliation{Joint Insitute for Nuclear Research, Dubna, Russia} 
\author{A.~D.~Kovalenko}
   \affiliation{Joint Insitute for Nuclear Research, Dubna, Russia}  
\author{M.~Haiduc}
   \affiliation{Institute of Space Sciences, Magurele, Romania}
 \author{A.~I.~Malakhov}
   \affiliation{Joint Insitute for Nuclear Research, Dubna, Russia} 
\author{G.~I.~Orlova}
   \affiliation{Lebedev Institute of Physics, Russian Academy of Sciences, Moscow, Russia} 
\author{P.~A.~Rukoyatkin}
   \affiliation{Joint Insitute for Nuclear Research, Dubna, Russia} 
\author{V.~V.~Rusakova}
   \affiliation{Joint Insitute for Nuclear Research, Dubna, Russia} 
\author{T.~V.~Shchedrina}
   \affiliation{Joint Insitute for Nuclear Research, Dubna, Russia}
\author{E.~Stan} 
   \affiliation{Institute of Space Sciences, Magurele, Romania}
\author{R.~Stanoeva}
  \affiliation{Institute for Nuclear Research and Nuclear Energy, Sofia, Bulgaria}
 \author{I.~Tsakov}
   \affiliation{Institute for Nuclear Research and Nuclear Energy, Sofia, Bulgaria}  
 \author{P.~I.~Zarubin}
     \email{zarubin@lhe.jinr.ru}    
     \homepage{http://becquerel.lhe.jinr.ru}
   \affiliation{Joint Insitute for Nuclear Research, Dubna, Russia} 
 \author{I.~G.~Zarubina}
   \affiliation{Joint Insitute for Nuclear Research, Dubna, Russia}   

\date{\today}

\begin{abstract}
\indent The results of investigations of the relativistic $^9$Be nucleus
 fragmentation in emulsion which entails the production of two He fragments
 of an energy of 1.2~A~GeV are presented. The results of the angular
 measurements of the $^9$Be$\rightarrow$2He events are analyzed.
 The $^9$Be$\rightarrow^8$Be+n fragmentation channel involving the $^8$Be decay
  from the ground (0$^+$) and the first excited (2$^+$) states to two
 $\alpha$ particles is observed to be predominant.\par
\end{abstract}

\maketitle
\section{\label{sec:level1}Introduction}
\indent The $^9$Be nucleus is a loosely bound n+$\alpha+\alpha$ system.
 The energy threshold of the $^9$Be$\rightarrow n+\alpha+\alpha$ dissociation channel
 is 1.57~MeV. Investigations of the $^9$Be fragmentation are of interest for
 astrophysics, in particular for the problems of nuclear synthesis of chemical
 elements with atomic number~A$>$8.\par
\indent The study of the $^9$Be fragmentation at relativistic energies gives
 the possibility of observing the reaction fragments which are the decay products
 of unbound $^8$Be and $^5$He nuclei\cite{Lepekhin}. The method of nuclear emulsions
 used in the present paper allows one to observe the charged component of the relativistic
 $^9$Be$\rightarrow$2He+n fragmentation channel. Owing to a good angular resolution of
 this method it is possible to separate the $^9$Be fragmentation events which are
 accompanied by the production of an unstable $^8$Be nucleus with its subsequent
 breakup to two $\alpha$ particles. In this case, the absence of a combinatorial
 background  (of three and more $\alpha$ particles) for $^9$Be which is typical of
 heavier N$\alpha$ nuclei  $^{12}$C and $^{16}$O  makes it possible to observe
 distinctly this picture. The present paper is of importance for estimating the
 $^8$Be role in more complicated N$\alpha$ systems.\par

\section{\label{sec:level2}Experiment}

\indent  Nuclear emulsions were exposed to relativistic $^9$Be nuclei at the JINR
 Nuclotron. A beam of relativistic $^9$Be nuclei was obtained in the $^{10}$B$\rightarrow^{9}$Be
 fragmentation reaction using a polyethylene target. The $^9$Be nuclei constituted about
 80\% of the beam , the remaining  20\%  fell on Li and He nuclei.\par
\begin{figure*}
    \includegraphics{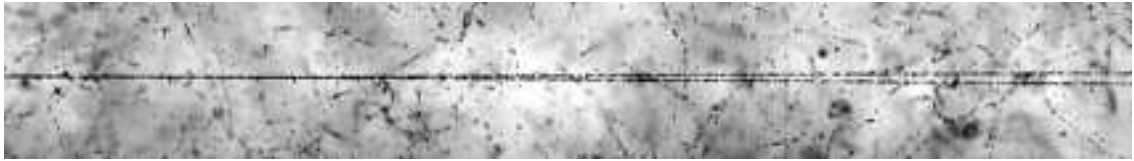}
    \caption{\label{fig:1} An event of the type of \lq\lq white\rq\rq star from the
     fragmentation of a relativistic $^9$Be nucleus into two He fragments in emulsion.
     The photograph was obtained on the PAVIKOM (FIAN) complex.}
    \end{figure*}

\indent The exposed emulsion stack consistent  of 15 layers of the BR-2 emulsion.
 The layer thickness and dimensions were 600 $\mu$m and 10$\times$20 cm$^2$, respectively.
 Events were sought by viewing the particle tracks by mean of the MBI-9 microscope.
 We found about 160 events of the $^9$Be fragmentation  involving the two He fragment
 production in the forward fragmentation cone with a polar angle of 6$^{\circ}$(0.1 rad).
The requirement of conservation of the fragment charge  in the fragmentation cone was
 fulfilled for the detected events. We allowed 5--7 tracks of various types in a wide
 (larger than 6$^{\circ}$) cone for the purpose of accumulating  additional statistics.
The charge of the He fragment tracks  was estimated by sight, for the emulsion method
  makes it possible to distinguish reliably the H and He isotopes. An example of the
  $^9$Be$\rightarrow$2He fragmentation event in emulsion is given in
 Fig.~\ref{fig:1} \cite{web}. This event belongs to the class of \lq\lq white\rq\rq
 stars  as far as it contains neither target nucleus fragments, nor produced mesons.\par
 \indent The angles of the tracks in emulsion for the detected events  were measured
 by the KSM-1 microscope. We measured  the coordinates of ten points on the primary
 nucleus track and of ten points on each of the fragment tracks. The points were
 selected to be spaced by a step of 100 $\mu$m, the overall track length used for
 measurement being 1 mm. By suggesting a linear dependence between the coordinates
 of the track points  the least square method was used to find  the p$_{0}$ and p$_{1}$
 coefficients of the first-degree approximating polynomial of the z(x) and y(x)
 coordinate dependences. The coordinates of a supposed interaction point (vertex) were
 suggested to be equal to zero. The angles were calculated by the found coefficients.
 At present angular measurements were carried out for 70 fragmentation events.\par
\begin{figure}
    \includegraphics[width=5in]{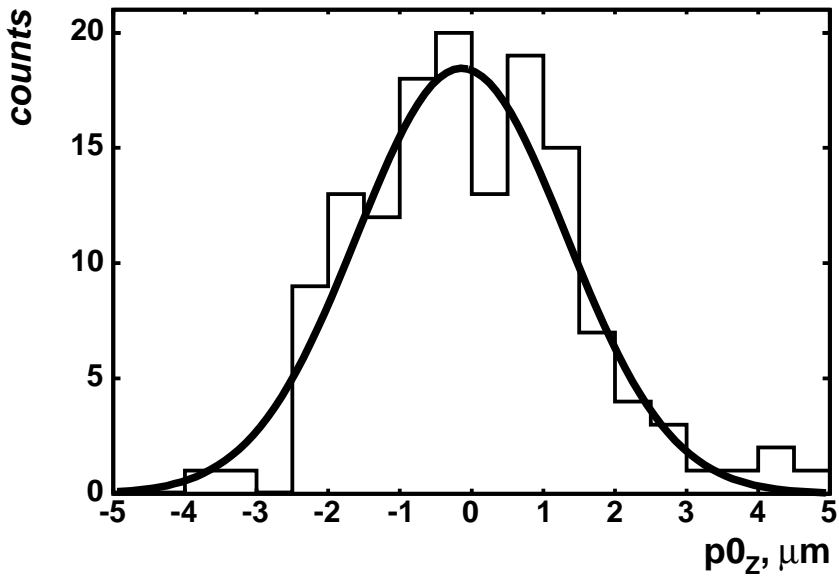}
    \caption{\label{fig:2} The  distribution  of  the p$_{0z}$ coefficients for the z(x)
 coordinate dependence for He fragment tracks.}
    \end{figure}

 \indent The accuracy of measurements of the He fragment emission angles was estimated
 on the basis of the p$_{0z}$ coefficient distribution for z(x)=p$_{0z}$+p$_{1z}\cdot$x
in experiment (Fig.~\ref{fig:2}). The p$_{0z}$ value shows a divergence  between
 the measured and calculated z coordinates of the event vertex. In this case, the
 z coordinate was measured less accurately which is explained by a specific treatment
 of the emulsion layers as well as by measurement errors. For example, due to treatment
 the emulsion layer thickness decreases by about a factor of two  and the error in z
 coordinate measurements depends on an exact focusing on the track. For the track length
  of 1 mm the measurement accuracy was found to be not worse than 4.5$\cdot10^{-3}$ rad.,
 which was about  34 MeV/c (see eq.\ref{eq1}) after evaluating it in the $\alpha$ particle
 transverse momenta.\par
 \indent The opening angle $\Theta$ of two He fragments was measured as the angle of
 one of the tracks with respect to the other. This way enabled us to determine the
 opening angle more accurately by decreasing  distortion effects in a layer. Thus in
 experiment a mean value of the measurement error was 1.3$\cdot10^{-3}$ rad., which was quite
 enough for separating events involving the $^8$Be production. A distinctive feature of
 the experiment using the emulsion method consists in that the quality of measurements
 of small $\Theta$ angles between tracks (4--6$\cdot10^{-3}$ rad.) depends on the conditions of treatment and
 storage of layers, as well as on the positioning of an event  in the layer.
 For example, distortions for small $\Theta$ are seen to depend on the mutual orientation
 of the plane of an emulsion layer and that of a track pair. The strongest distortions
 are observed when the planes are perpendicular to each other which affects the shape
 of the angular distribution for $\Theta$(Fig.~\ref{fig:4}) by approaching it to zero.\par
 
 \begin{figure}
    \includegraphics[width=7in]{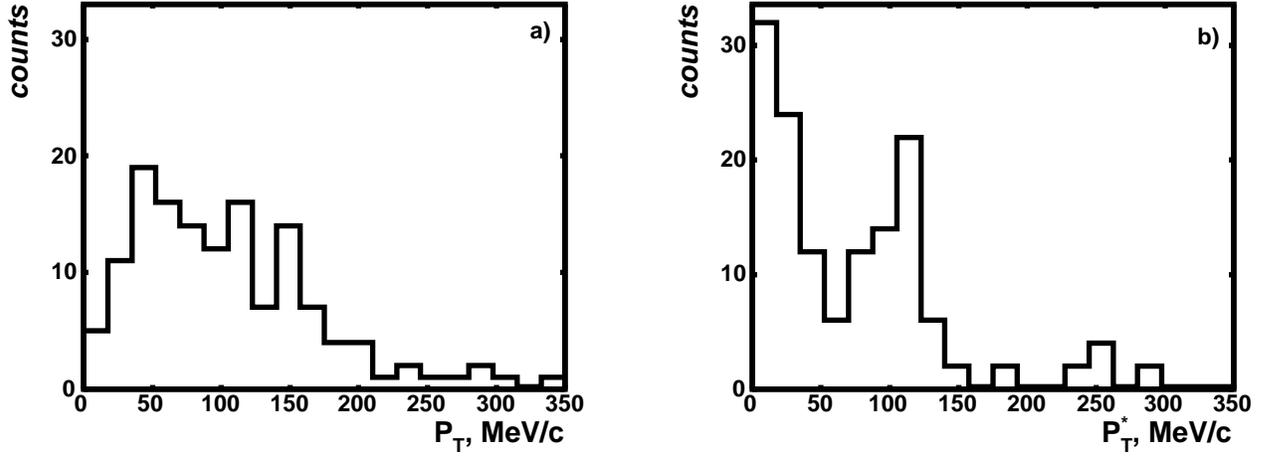}
    \caption{\label{fig:3} The P$_{T}$ transverse momentum distribution  of $\alpha$ particles
 in the laboratory system (a), and the P$^*_{T}$ momentum distribution in the c.m.s. of an
 $\alpha$ particle pair (b).}
    \end{figure}
\section{\label{sec:level3}Results}
\indent In analyzing the data  both He fragments observed in the $^9$Be$\rightarrow$2He+n
 channel were supposed to be $\alpha$ particles. This assumption is motivated
  by the fact that at small angles  the $^9$Be$\rightarrow$2$^4$He+n fragmentation
 channel with an energy threshold of 1.57~MeV must dominate the $^9$Be$\rightarrow^3$He+$^4$He+n
 channel whose energy threshold is 22.15~MeV. The $^3$He fraction will not exceed a
 few percent in this energy range \cite{Belaga96} and all the He fragments in the
 detected events may be thought of as $\alpha$ particles.\par
\indent In Fig.~\ref{fig:3}a  the P$_{T}$ transverse momentum distribution  of $\alpha$
 particles in the laboratory frame of reference is calculated without the account of
 particle energy losses in emulsion  by the equation    
\begin{equation}
{P_{T}=p_{0}\cdot A\cdot sin\theta }\label{eq1}
\end{equation}
where p$_{0}$, A and $\theta$ are the momentum per nucleon, the fragment mass
 and  the polar emission angle, respectively.  The mean value of the transverse momentum
 in the laboratory system was $<P_{T}>$=109 MeV/c, and the distribution FWHM was
 $\sigma$=66 MeV/c. This may be an indication of the fact that the experimental data are not
 of the same kind which can be pronounced when going over to the c.m.s. of two $\alpha$ particles.\par
 \indent  The P$^*_{t}$ transverse momentum distribution of $\alpha$ particles in the c.m.s.
 of two $\alpha$ particles described by the equation
\begin{equation}
   {\bf P}^{*}_{Ti}\cong {\bf P}_{Ti}- \frac{\sum_{i=1}^{n}{\bf P}_{Ti}}{n_{\alpha}}\label{eq2} 
 \end{equation}
where P$_{Ti}$ is the transverse momentum  of an i-th $\alpha$ particle in the laboratory
 system n$_{\alpha}$=2  is given in Fig.~\ref{fig:3}b.  There is observed a grouping of
 events around  two peaks with the values $<P^*_{Ti}>\approx$25~MeV/c and $<P^*_{Ti}>\approx$102~MeV/c.
 In \cite{Avetyan96} the appropriate mean values of the $\alpha$ fragment transverse momenta
 are  $<P^*_{Ti}>\approx$121 MeV/c for $^{16}$O$\rightarrow$4$\alpha$,
 $<P^*_{Ti}>\approx$141 MeV/c \cite{Belaga95} for $^{12}$C$\rightarrow$3$\alpha$
 and $<P^*_{Ti}>\approx$200 MeV/  for $^{22}$Ne$\rightarrow$5$\alpha$
 (processing of the available data). There by we clearly see a tendency toward an increase
 of the mean $\alpha$ particle momentum with increasing their multiplicity. This implies a
 growth  of the total coulomb interaction of alpha clusters arising in nuclei.\par
\begin{figure}
    \includegraphics[width=5in]{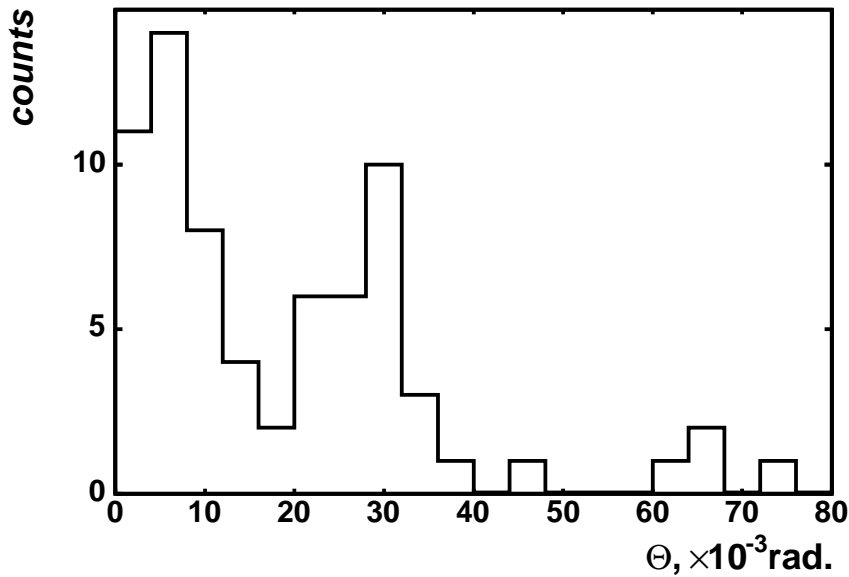}
    \caption{\label{fig:4}  The opening $\Theta$ angle distribution of $\alpha$ particles
 in the $^9$Be$\rightarrow$2$\alpha$ fragmentation reaction at 1.2~A~GeV energy.}
    \end{figure}
\indent  A particular feature of the observed $^9$Be$\rightarrow$2$\alpha$ fragmentation
 is a total deflection of an $\alpha$ particle pair with respect to the axis associated
 with the primary nucleus direction.  The average value of  a \lq\lq missing\rq\rq transverse
 momentum P$_{Tmiss}$ for 28 events of the type of  the \lq\lq white\rq\rq star is
 $<P_{Tmiss}>\approx$127 MeV/c.  This effect can be explained both by the influence
 of a neutron which is \lq\lq invisible\rq\rq in emulsion and by the recoil nuclei.
 The $^8$Be production channel is not the only possible one which involves the two He
 fragment production. In particular, in \cite{Fulton} one discusses the possibility of
 observing  a channel  $^9$Be$\rightarrow^5$He+$\alpha\rightarrow$2$\alpha$+n.
 In the present paper, this channel is not considered  as far as it is impossible
 to observe a neutron (the latter leaves no track in emulsion).\par   
\indent In the opening angle $\Theta$ distribution (Fig.~\ref{fig:4}) one can also see
 two peaks with mean values 4.5$\cdot 10^{3}$rad. and 27$\cdot 10^{-3}$rad.
 The ratio of the numbers of the events in the peaks  is close to unity.\par
 \indent  The $|Theta$ distribution entails the invariant energy  Q$_{2\alpha}$
 distribution which is calculated as a difference between the effective invariant mass
  M$_{2\alpha}$ of an $\alpha$ fragment pair and the doubled  $\alpha$ particle mass by
 the equation
\begin{eqnarray}
   {M^2_{2\alpha}=-(\sum_{j=1}^{2} P_j)^2} \nonumber\\
   {Q_{2\alpha}=M_{2\alpha} - 2\cdot m_{\alpha}}
 \end{eqnarray}
where P$_j$ is the $\alpha$ particle 4-momentum.\par
\indent In the invariant energy Q$_{2\alpha}$ distribution there are two peaks in the ranges
  0 to 1 MeV and 2 to 4 MeV. The shape of the distribution does not contradict the suggestion
 about the $^9$Be fragmentation involving the production of an unstable $^8$Be nucleus which
 decays in the 0$^+$ and 2$^+$ states. The values of the peaks of the invariant energy
 Q$_{2\alpha}$ and the transverse momenta  P$^{*}_{T}$ in the c.m.s. relate to each other.
 To the Q$_{2\alpha}$ range from 0 to 1~MeV with a peak at 100~keV there corresponds a peak
  P$^*_{T}$ with  $<P^*_{Ti}>\approx$25~MeV/c , and to the Q$_{2\alpha}$ range from
 2 to 4~MeV  there corresponds a peak with $<P^*_{Ti}>\approx$102~MeV/c.\par
\begin{figure}
    \includegraphics[width=5in]{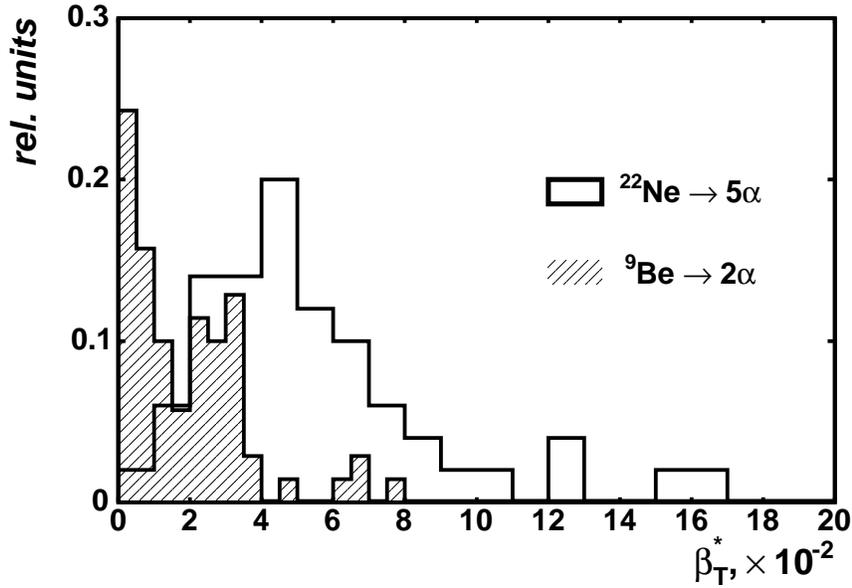}
    \caption{\label{fig:6}  The $\beta_{T}$ velocity distribution in the c.m.s. of $\alpha$ 
particles  in the fragmentation process $^9$Be$\rightarrow$2$\alpha$ at 1.2~A~GeV and in
 the fragmentation process $^{22}$Ne$\rightarrow$5$\alpha$ at 3.7 A GeV.   
}
    \end{figure}

 \indent  Fig.~\ref{fig:6} gives the $\beta^*_ {T}$ velocity distribution of $\alpha$
 particles in their c.m.s. which were produced in the 1.2~A~GeV $^9$Be$\rightarrow$2$\alpha$
 fragmentation process as compared to the 3.7~A~GeV $^{22}$Ne$\rightarrow$5$\alpha$ process.
 In both cases the velocity values are in a non-relativistic domain. The distribution for the
 $^{22}$Ne nucleus is essentially wider, its larger average value  reflects the increase in
 the $\alpha$ particle transverse momenta. Thus the study of the relativistic $^9$Be
 fragmentation in emulsion will make it possible in future to employ the obtained data in
 the analysis of the angular distributions of more complicated systems.\par 

\section{\label{sec:level3}Conclusion}
\indent We summarize the results of the study of the relativistic $^9$Be fragmentation
 in emulsion. 160 events of the $^9$Be$\rightarrow$2He fragmentation were found.
 Angular measurements for 70 events were carried out  with an accuracy  not worse
 than 4.5$\cdot 10^{-3}$rad. The results were used to obtain the average value of the
 $\alpha$ particle transverse momenta  $<P_{T}>\approx$109~MeV/c  in the laboratory frame
 of reference.\par
 \indent  When going over to the c.m.s. of two $\alpha$ particles, in the $P^*_{T}$ momentum
 distribution of $^4$He fragments there is observed the formation of two peaks with the mean
 values $<P^*_{Ti}>\approx$25~MeV/c and $<P^*_{Ti}>\approx$102~MeV/c which is in agreement
 with the suggestion about the $^9$Be fragmentation involving the $^8$Be production.\par
\indent  In the Q$_{2\alpha}$ invariant energy distribution of an $\alpha$ particle pair
 there is observed a separation of virtually all the events over the two energy intervals:
 from 0 to 1 MeV with a peak at 100 keV and from 2 to 4 MeV.  This fact suggests the dominance
 of the $^9$Be$\rightarrow ^8$Be+n fragmentation accompanied by  a $^8$Be decay from the
 ground (0$^+$) and the first excited (2$^+$) states  to two $\alpha$ particles.\par
\indent The data obtained from $^9$Be angular measurements  can be employed for the
 estimation of the role of $^8$Be in more complicated N$\alpha$ systems.\par
\begin{acknowledgments}
\indent The work was supported by the Russian Foundation for Basic Research
 ( Grants 96-159623, 02-02-164-12a,03-02-16134, 03-02-17079 and 04-02-16593 ),
 VEGA 1/9036/02.  Grant from the Agency for Science of the Ministry for Education of the
 Slovak Republic and the Slovak Academy of Sciences, and Grants from the JINR
 Plenipotentiaries of the Republic of Bulgaria, the Slovak Republic, the Czech Republic
 and Romania in the years 2002-2005.\par
 \indent  In conclusion, the authors are indebted to the JINR collaborators  A.M. Sosulnikova,
 I.I. Sosulnikova and G.V. Stelmakh who performed visual searches for events. Our special
 thanks are to F.G. Lepekhin (St.Petersburg Institute of Nuclear Physics) for useful methodical
 recommendations concerning the treatment procedure.\par  
\end{acknowledgments}    
    
\begin{figure}
    \includegraphics[width=5in]{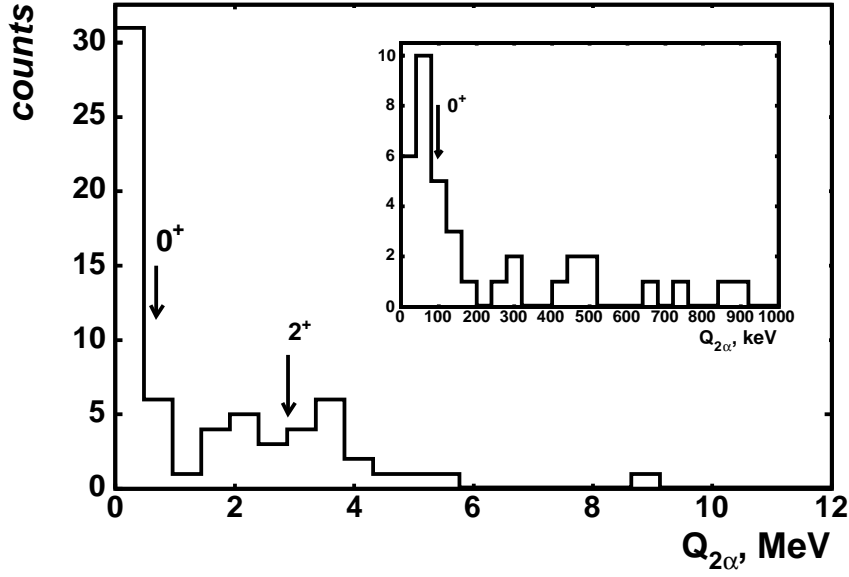}
    \caption{\label{fig:5} The invariant energy Q$_{2\alpha}$ distribution of $\alpha$
 particle pairs in the $^9$Be$\rightarrow$2$\alpha$ fragmentation reaction at 1.2~A~GeV energy.
  On the intersection: the Q$_{2\alpha}$ range  from 0 to 1~MeV.
 Arrows show the location of the $^8$Be 0$^+$ and 2$^+$ nucleus levels.}
    \end{figure}
    
\newpage


\begin{thebibliography}{}
\bibitem{Lepekhin}
F. G. Lepekhin et al., Phys. Letters , \textbf{58}, 493-496 (1993).
\bibitem{web}
Web site of the BECQUEREL Project, http://becquerel.jinr.ru/
\bibitem{Belaga96}
V. V. Belaga et al., Phys. At. Nucl., \textbf{59}, 869-877 (1996).
\bibitem{Avetyan96}
F. A. Avetyan et al., Phys. At. Nucl., \textbf{59}, 110-116 (1996).
\bibitem{Belaga95}
V. V. Belaga et al., Phys. At. Nucl., \textbf{58}, 2014-2020 (1995).
\bibitem{Fulton}
B. R. Fulton et al., Phys. Rev., C \textbf{70} 047602(2004).
\end{thebibliography}
\end{document}